\documentclass[pre,preprint,a4paper,superscriptaddress]{revtex4}
\usepackage{epsf}
\usepackage{amssymb}
\usepackage{graphicx}% Include figure files
\usepackage{epsfig}
\newcommand{\bd     }{\begin{displaymath}}
\newcommand{\ed     }{\end{displaymath}}

\newcommand{\order  }{{\cal O}}

\newcommand{\s      }{\sigma}

\newcommand{\bra    }{\langle}
\newcommand{\ket    }{\rangle}
\newcommand{\Bra    }{\left\langle}
\newcommand{\Ket    }{\right\rangle}

\newcommand{\bJ     }{\mbox{\boldmath$J$}}
\newcommand{\bc     }{\mbox{\boldmath$c$}}

\newcommand{\bsigma }{{\mbox{\boldmath$\sigma$}}}

\newcommand{\btau   }{{\mbox{\boldmath$\tau$}}}

\newcommand{\rme    }{{\mathrm{e}}}

\newcommand{\rmd    }{\textrm{d}}
\newcommand{\rmi    }{\textrm{i}}

\begin{document}

\title{\bf Dynamic rewiring in small world networks}

\author{J P L Hatchett}\affiliation{Laboratory for Mathematical Neuroscience, RIKEN Brain Science Institute, Saitama 351-0198, Japan}
\email{hatchett@brain.riken.jp}
\author{N S Skantzos}\affiliation{Instituut voor  Theoretische Fysica, Celestijnenlaan 200D, K.U.Leuven B-3001, Belgium}
\email{Nikos.Skantzos@fys.kuleuven.be}
\author{T Nikoletopoulos}\affiliation{Department of Mathematics,
King's College London, Strand WC2R 2LS,
U.K.}\email{theodore@mth.kcl.ac.uk}

\begin{abstract}
We investigate equilibrium properties of small world networks, in
which both connectivity and spin variables are dynamic, using
replicated transfer matrices within the replica symmetric
approximation. Population dynamics techniques allow us to examine
order parameters of our system at total equilibrium, probing both
spin- and graph-statistics. Of these, interestingly, the degree
distribution is found to acquire a Poisson-like form (both within
and outside the ordered phase). Comparison with Glauber simulations
confirms our results satisfactorily.
\end{abstract}

\pacs{75.10.Nr, 05.20.-y, 89.75.-k} %64.60.Cn
\preprint{FoG 2005/03}

\maketitle

%%%%%%%%%%%%%%%%%%%%%%%%%%%%%%%%%%%%%%%%%%%%%%%%%%%%%%%%%%%
\section{Introduction}
%%%%%%%%%%%%%%%%%%%%%%%%%%%%%%%%%%%%%%%%%%%%%%%%%%%%%%%%%%%

Small worlds are systems characterized by a local neighborhood
(given by short-range bonds) with a sparse set of long-range
connections per spin. This simple architectural effect has been
shown to bring about remarkable cooperative and synchronization
phenomena. The term `small-world' has been coined by the now famous
experiment by the Harvard social psychologist Stanley Milgram
\cite{milgram}: in 1967, as part of his research on the network of
acquaintances in the United States, he took a number of letters and
handed them over to people totally unrelated with the addressees,
and, with the instructions to pass them over to someone they think
might know the addressee. This process was repeated until the
letters finally arrived to their destination. Stanley Milgram then
estimated the average path length from the two randomly chosen
individuals which turned out to be a mere six. This experiment
revealed that although social networks are very sparse, in reality
any two pair of nodes can be topologically very close. In fact,
numerical studies of other types of real networks (e.g.\@ citation,
linguistic, disease spreading, etc) show that the small-world effect
is a common architecture among real network structures and brings
about optimal information processing. The question then arises, how
do networks spontaneously evolve from (almost) random configurations
into particular structures such as small-world ones? And which
underlying process drives the distribution of the long-range
short-cuts within the nodes? The above questions fall under a
particularly active area of research, namely the evolution of
networks (see e.g.\@ \cite{Chan03,barrat}, or \cite{doro}, for a recent review). 
Since real networks (be it
biological, social, economic or otherwise) hardly ever maintain a
static architecture this problem of predicting network structure has
important applications. In this paper we attempt to formulate and
describe the thermodynamics of the problem from an analytic point of
view. This carries the obvious set of advantages and disadvantages:
while resulting in robust and exact results, it will be amenable to
a set of (perhaps not fully realistic) assumptions. To be precise,
we examine a coupled system on a small-world architecture in which
both nodes and connections are mobile. However, the two dynamic
processes occur on distinct timescales; connections are assumed to
evolve slowly enough such that, at each of their update steps, spins
have effectively reached equilibrium. This will allow us to avoid
solving the explicit dynamical relations and instead turn directly
to the thermodynamics. Our starting point is the free energy per
connection degree of freedom. We couple the two dynamic processes of
the spins and the connections by constructing two Hamiltonians: a
typical Ising one describing the energy of the spins and a
Hamiltonian of the connections, constructed to reward network
configurations minimizing the free energy of the spins. This choice
allows us to proceed analytically while retaining a sufficient
amount of realism. The result is a replica theory where the replica
dimension represents the ratio between the two temperatures (of the
spins vs connection processes).

Our paper is organised as follows: In the following section we introduce our model and
the pair of energy functions describing the thermodynamics of the spins and the graph variables.
In section \ref{sec:f_extr} we write the total free energy of the system as an extremisation
problem in terms of the typical finite-connectivity order parameter function. We then proceed
to define the observables of our system, of which there are here two kinds, probing spin (section \ref{sec:Spin_system_observables}) and graph (section \ref{sec:connectivity_observables}) organization statistics respectively.
The single pure state approximation (section \ref{sec:RS}) allows us to deal with the resulting replicated transfer matrices
following the diagonalization process of \cite{diagonalisation,guzaiSW}.
We first derive in section \ref{sec:Self-consistent_order_function_equations} numerically tractable forms for our set of order functions which are to be solved via population
dynamics. Observables such as magnetization, average connectivity, or degree distribution then follow easily, see section \ref{sec:Observables within replica symmetry}.
We perform a bifurcation analysis and plot phase diagrams showing the transition lines
between ordered and paramagnetic phases in section \ref{sec:phase_diagrams}. We find that, perhaps contrary to initial expectations, the resulting degree distributions
are close to, or exactly, Poisson. Comparison with numerical
simulations shows good agreement given the complexity of these experiments
requiring adiabatic (practically infinitely long) timescales.

%%%%%%%%%%%%%%%%%%%%%%%%%%%%%%%%%%%%%%%%%%%%%%%%%%%%%%%%%%%
\section{Model description}
%%%%%%%%%%%%%%%%%%%%%%%%%%%%%%%%%%%%%%%%%%%%%%%%%%%%%%%%%%%
We study a system of $N$ Ising spins
$\bsigma=(\sigma_1,\ldots,\sigma_N)$ with $\sigma_i\in\{-1,1\}$,
arranged on a ``small-world'' structure. We represent this by a
one-dimensional lattice with uniform nearest-neighbor interactions
of strength $J_0$ and with randomly-chosen sparse short-cuts of
strength $J_{ij}\in\{-J,J\}$ that can connect distant pairs of spins
$(i,j)$.  For every $i\neq j$ we assign a variable $c_{ij}$ denoting
whether a connection exists ($c_{ij}=1$) or not ($c_{ij}=0$), with
$c_{ii}=0$. In the absence of short-cuts the average path length is
$N/4$ while in the combined system the scaling is bounded above by
$\log(N)$. This significant reduction in the path length is commonly
termed the ``small-world'' effect \cite{watts-strogatz}. For static architectures, in
which the link and bond matrices $\{c_{ij},J_{ij}\}$ are taken as
quenched random variables, frustration effects are known to induce
spin-glass phases \cite{guzaiSW}. Our model aims to examine
thermodynamic properties of the above spin systems under the freedom
of allowing the connectivity and bond matrices $\{c_{ij},J_{ij}\}$
to evolve in time in search of the state that best promotes order
within the system. To be precise, on short timescales the links and
bonds can be seen as static variables with respect to which the
spins equilibrate, while on longer timescales $c_{ij}$ and $J_{ij}$
explore their configuration space. The measure of this latter
process is related to the ordering within the spin system on the
instantaneous state of the graph. Thus the spins and the graph
architecture on which they live are dynamically interwoven. It is
quite natural that the architecture dynamics depends on the entire
system state (including the spins) rather than just the architecture
itself (as with e.g.\@ preferential attachment \cite{barabasi}).
Links and connectivities are taken here to evolve on identical
timescales, although generalizing this to more involved scenarios is
also possible. On the timescale in which spins reach thermal
equilibrium our combined system is described by the ``fast''
Hamiltonian
\begin{equation}
H_f(\bsigma,\bc,\bJ) = -J_0 \sum_i \sigma_i \sigma_{i+1} - \sum_{i \leq   j} \sigma_i J_{ij} c_{ij} \sigma_j
\label{eq:Hf}
\end{equation}
(where we take periodic boundary conditions on the chain). Spins
equilibrate with respect to (\ref{eq:Hf}) at an inverse temperature
$1/\beta_f$, and their behavior is described by the partition
function
\begin{equation}
Z_f(\bc,\bJ) = \sum_{\bsigma} \rme^{-\beta_f H_f(\bsigma,\bc,\bJ)}
\end{equation}
On timescales sufficiently long to guarantee that spins have reached
equilibrium, links and bonds are not static, but evolve dynamically
and we will take their stationary state to be described by the
``slow'' Hamiltonian
\begin{eqnarray}
H_s(\bc,\bJ) &=& -\frac{1}{\beta_f} \log Z_f(\bc,\bJ) + V(\bc,\bJ)
\label{eq:Hs}
\\
V(\bc,\bJ) & = & \frac{1}{\beta_s}\sum_{i<j}c_{ij}\left[\log\left(\frac{N}{c}\right)
+\log\cosh(K_p )-K_p \frac{J_{ij}}{J}\right]
\label{eq:chemical}
\end{eqnarray}
This choice energetically favors those configurations of
$\{c_{ij},J_{ij}\}$ that minimize the free energy of the spins. The
role of the chemical potential $V(\bc,\bJ)$ is twofold: firstly, it
aims to preserve the overall nature of the small-world system; it
guarantees that for $N\to\infty$ the average number of connections
per spin is a finite number. Secondly, it allows us to tune the
relative concentration of $\{-J,J\}$ bonds in the system (as we will
see in section \ref{sec:connectivity_observables}, the former is
controlled by the variable $c$ whereas the latter by $K_p$). The
connectivity- and bond-variables $\{c_{ij},J_{ij}\}$ equilibrate
with respect to this slow Hamiltonian at inverse temperature
$\beta_s$, leading to a total partition function
\begin{equation}
Z_s = \sum_{\bc,\bJ} \rme^{-\beta_s H_s(\bc,\bJ)} = \sum_{\bc,\bJ}
[Z_f(\bc,\bJ)]^{\beta_s/\beta_f}\ \rme^{-\beta_s V(\bc,\bJ)}
\label{eq:slow_partition}
\end{equation}
This partition function, by construction, contains $n =
\beta_s/\beta_f$ replicas of the fast system. In general, the ratio
of inverse temperatures $n$ can take any value (integer or
otherwise) so that analytic continuation in the replica dimension
depends solely on our choice of temperature values. The limit $n\to
0$ corresponds to temperatures $T_s\to \infty$ in which the
partition sum (\ref{eq:slow_partition}) is dominated by the entropy
of the slow system. In contrast, $T_s\to 0$ favors those
architectures $\{c_{ij},J_{ij}\}$ that increase order among the spin
variables for a given number of links. Note that this is a general
optimization criterion which does not enforce \emph{a priori} any
particular structure on the links but allows the links to arrange
themselves. In fact, the graph statistics become interesting
observables, which we can measure, rather than enforced constraints.
Our order parameters follow from the slow free energy per spin
\begin{equation}
f_s = -\lim_{N \to \infty} \frac{1}{\beta_s N} \log Z_s
\label{eq:fs}
\end{equation}
and derivatives of this generating function.

%%%%%%%%%%%%%%%%%%%%%%%%%%%%%%%%%%%%%%%%%%%%%%%%%%%%%%%%%%%%
\section{The free energy}
\label{sec:f_extr}
%%%%%%%%%%%%%%%%%%%%%%%%%%%%%%%%%%%%%%%%%%%%%%%%%%%%%%%%%%%%

To calculate the slow partition function (\ref{eq:slow_partition})
we first take the trace over the connectivity and bond variables
$\{c_{ij}, J_{ij}\}$
\begin{equation}
Z_s = \sum_{\bsigma_1\ldots\bsigma_N}\rme^{\beta_f J_0 \sum_i
  \bsigma_i \cdotp \bsigma_{i+1}} \prod_{i < j} \left[1 +
  \frac{c}{N} \left\langle\rme^{\beta_f J \bsigma_i \cdotp \bsigma_j}
  \right\rangle_J  \right]
\end{equation}
up to irrelevant multiplicative constants. We denote $\bsigma_i =
(\sigma_i^1,\ldots, \sigma_i^n)$ where
$\bsigma_i\cdot\bsigma_j=\sum_\alpha\s_i^\alpha\s_j^\alpha$ and
defined the abbreviation $\Bra
f(J)\Ket_J=[2\cosh(K_p)]^{-1}[e^{K_p}f(J)+e^{-K_p}f(-J)]$. We are
interested in the case where $c$ (the chemical potential for bonds)
is finite, and hence $c/N\to 0$ in the limit $N\to\infty$, so that
the above product can alternatively  be seen as a product over
exponentials (up to terms of $\mathcal{O}(N^{-2})$). We thus
encounter the typical nested exponential form of finite connectivity
problems. To achieve site factorization it is convenient to
introduce into our expressions the order parameter function
\cite{monasson98,FoG0}
\begin{equation} P(\bsigma) = \frac{1}{N} \sum_i
\delta_{\bsigma,\bsigma_i} \label{eq:spin_distribution}
\end{equation}
via appropriately defined delta functions, which is a probability
distribution over replicated spins. In the limit $N\to\infty$ we can
now evaluate the free energy (\ref{eq:fs}) via steepest descent and
express it as an extremization problem in the space of probability
distributions $P(\bsigma)$, namely:
\begin{eqnarray}
 f_s &=& \mbox{extr}_{\{P(\bsigma)\}} \Bigg\{ \frac{c}{2\beta_s}
 \sum_{\bsigma \bsigma^\prime} P(\bsigma)
P(\bsigma^\prime) \Bra \rme^{\beta_f J \bsigma \cdotp
  \bsigma^\prime}\Ket_J\nonumber
- \lim_{N\to\infty}\frac{1}{\beta_s N} \log \sum_{\bsigma_1 \ldots
\bsigma_N} \prod_i T_{\bsigma_i,\bsigma_{i+1}}[P]\Bigg\}
\label{eq:fs_extr}
\end{eqnarray}
where $T_{\bsigma,\bsigma'}[P]$ represent the transfer matrix
elements
\begin{eqnarray}
T_{\bsigma, \bsigma^\prime}[P] = \exp\left[ \beta_f J_0 \bsigma
\cdotp
  \bsigma^\prime + c \sum_\btau P(\btau) \Bra \rme^{\beta_f J \bsigma
  \cdotp \btau} \Ket_J \right]\label{eq:transfer_matrix}
\end{eqnarray}
and $P(\bsigma)$ is to be evaluated from the fixed-point equation
\begin{equation}
P(\bsigma)=\frac{{\rm Tr}\left(Q[\bsigma]\,T^N[P]\right)}{{\rm
Tr}\left(T^N[P]\right)} \label{eq:Ptr} \hspace{10mm}
Q_{\bsigma,\bsigma'}[\btau]\equiv\delta_{\bsigma,\btau}\delta_{\bsigma,\bsigma'}
\end{equation}
For more details on the derivation of the above expressions we refer
the reader to \cite{guzaiSW} where the special case of the $n\to 0$
limit was studied.

Finding solutions of (\ref{eq:Ptr}) amounts to diagonalising the
transfer matrix $T$ of dimensionality $2^n\times 2^n$. This
problem has been solved in \cite{diagonalisation}. Here we will
not be concerned in the entire spectrum of eigenvalues, as the
limit $N\to\infty$ ensures that only the largest eigenvalue
$\lambda_0$ will provide a non-vanishing contribution to the free
energy. The left- and right- eigenvectors associated with this
eigenvalue follow from the equations
\begin{eqnarray}
\sum_{\bsigma'}T_{\bsigma,\bsigma'}[P]\ U(\bsigma')=\lambda_0\
U(\bsigma)\label{eq:T_left}\\ \sum_{\bsigma'}V(\bsigma')\
T_{\bsigma',\bsigma}[P]\ =\lambda_0\ V(\bsigma)\label{eq:T_right}
\end{eqnarray}
These eigenvectors are unique up to the usual arbitrary
multiplicative factor, and non-negative \cite{diagonalisation,
FoGs}. We note that we need both left- and right-eigenvectors since
the transfer matrix $T[P]$ is non-symmetric. The order function
$P(\bsigma)$ (\ref{eq:spin_distribution}) is manifestly normalized.
Due to our scaling freedom for the eigenvectors we can always choose
them so that $\sum_{\bsigma}U(\bsigma)=\sum_{\bsigma}V(\bsigma)=1$.
The physics of our system is given by the normalized distributions
$P(\bsigma)$, $V(\bsigma)$, $U(\bsigma)$ which are to be found by
self-consistently solving equations
(\ref{eq:transfer_matrix}-\ref{eq:T_right}) (in fact, $U(\bsigma)$
and $V(\bsigma)$ turn out to represent the  distributions of cavity
spins with a chain bond rather than a graph bond removed
\cite{FoGs}).

%%%%%%%%%%%%%%%%%%%%%%%%%%%%%%%%%%%%%%%%%%%%%%%%%%%%%%%%%%%%%%%%%%%%%%%%%%
\section{Observables}
%%%%%%%%%%%%%%%%%%%%%%%%%%%%%%%%%%%%%%%%%%%%%%%%%%%%%%%%%%%%%%%%%%%%%%%%%%
\label{sec:observables}

\subsection{Spin system observables}
\label{sec:Spin_system_observables}

We are interested in probing organizational properties of our
system both within the spin variables and the connectivity ones.
For the spin system, we define the canonical observables; the
magnetization and the overlap order parameter as moments of the
probability distribution (\ref{eq:spin_distribution}), namely
\begin{eqnarray}
m_\alpha&=&\sum_{\bsigma}P(\bsigma)\
\sigma^\alpha\label{eq:m_def}\\
q_{\alpha\beta}&=&\sum_{\bsigma}P(\bsigma)\
\sigma^\alpha\sigma^\beta \label{eq:q_def}
\end{eqnarray}
In the above and henceforth, the quantities $P(\bsigma)$,
$V(\bsigma)$, $U(\bsigma)$ are given by their saddle-point values.

It is well known that infinite dimensional systems, such as small
world lattices, with frozen bonds of random signs, will have a
spin-glass ground state at low temperatures for certain values of the control
parameters \cite{viana-bray85,
kanter-sompo87, mezard-parisi87}. This spin glass ordering is intimately linked to
frustration within the system; the inability of spins to find
energetically optimal configurations. By allowing the architecture
some limited degree of freedom, we expect that the system will be
able to optimize its state somewhat better. Probing the degree of
frustration within the system as the slow temperature is varied is
therefore an interesting problem. The frustration is normally
defined as the fraction of closed loops from sites $i_1 \to i_2 \to
\ldots i_k \to i_1$ where the product $J_{i_1i_2} \ldots J_{i_k
i_1}$ is negative. Unfortunately, to measure this directly in our
system where bonds are mobile would require us to be able to measure
correlations over long length scales within the system (in fact
scaling like the average loop length $\sim \log(N)$), which is
technically difficult. To try and finesse this problem, in
\cite{wemmenhove-skantzos-coolen, wemmenhove-skantzos}, the fraction
of misaligned spins was calculated, i.e. the fraction of spins that
did not point in the direction of their local field. Due to the
mobility of the connections in our system we expect that thermal
equilibrium states within the ordered phases will be steered towards
configurations where spin alignment with their local fields is
optimal. The result of this structural organization can be measured
by the following quantity
%\begin{equation}
$\phi=\int_{-\infty}^{0^-}\rmd h\,P(1,h)+\int_{0^+}^\infty \rmd h
\,P(-1,h)$
%\end{equation}
which gives the fraction of misaligned spins and is defined in
terms of the joint spin-field distribution
%\begin{equation}
$P(\sigma,h)=\lim_{N\to\infty}\frac{1}{N}\sum_i\bra\delta_{\sigma,\sigma_i}\delta[h-h_i(\bsigma)]\ket_{\rm s}$
%\end{equation}
where $\bra \cdots\ket_{\rm s}$ denotes thermal averages over the
slow process $\bra x\ket_{\rm s}=Z_s^{-1}\sum_{\bc,\bJ}e^{-\beta_s
H_s(\bc,\bJ)}x$ and $h_i(\bsigma)\equiv\sum_j
c_{ij}J_{ij}\s_j+J_0(\sigma_{i+1}+\sigma_{i-1})$ denotes the local
field at site $i$. However, at e.g. very low temperatures, one
expects the spins to align to their local fields whether they are in
a spin glass phase or not. Thus to try and get a different measure
to probe the frustration in the system we consider the fraction of
bonds in the graph, which are not energetically optimized by the
spin configuration:
\begin{eqnarray}
\psi = \frac1N \sum_i \Bra \Theta(-\sigma_i \sigma_{i+1} J_0) \Ket_{\rm s}
+ \frac{1}{cN} \sum_{i < j} \bra c_{ij} \Theta(-\sigma_i \sigma_{j}
J_{ij} ) \ket_{\rm s}
\end{eqnarray}
This is also not an absolute measure of frustration, but in the low
temperature spin glass phase $\psi$ will be non-zero, as opposed to
a low temperature ferromagnet where we would have $\psi = 0$. The
calculation of either $\psi$ or $\phi$ is similar to the calculation
of the free-energy, with a specific observable (i.e. matrix in the
transfer matrix notation) at one or two sites. We find
\begin{eqnarray}
\psi = D_1 \sum_{\bsigma \bsigma^\prime} V(\bsigma) \Theta(-\sigma_1
\sigma_1^\prime J_0) T_{\bsigma \bsigma^\prime}[P] U(\bsigma^\prime)
+ D_2 \sum_{\bsigma \bsigma^\prime} P(\bsigma) P(\bsigma^\prime)
\bra \Theta(-\sigma_1 \sigma_1^\prime J) \rme^{\beta_f J\bsigma
\cdotp \bsigma^\prime} \ket_J
\end{eqnarray}
where $D_1$ and $D_2$ are normalization constants to give the
fraction of sites, i.e.\@ $D_1 =\sum_{\bsigma \bsigma^\prime}
V(\bsigma) T_{\bsigma \bsigma^\prime}[P] U(\bsigma^\prime)$ and
$D_2=\sum_{\bsigma \bsigma^\prime} P(\bsigma)\,P(\bsigma')\Bra e^{\beta_f J\bsigma\cdot\bsigma'}\Ket$.

\subsection{Connectivity system observables}
\label{sec:connectivity_observables}

Let us now inspect organisational phenomena within the graph. We
first identify the roles played by the control parameters $c$ and
$K_p$ that appear in the chemical potential (\ref{eq:chemical}).
This can be done by adding suitable generating terms into the
Hamiltonian (\ref{eq:Hs}) and monitoring their impact on
(\ref{eq:fs_extr}). For instance, if one transforms $H_s\to
H_s+\lambda\frac{1}{c}\sum_{i<j}c_{ij}$ then taking derivatives
$\frac{\partial f_s}{\partial \lambda}|_{\lambda=0}$ translates to
\begin{equation}
\overline{c}\equiv \frac{1}{N}\sum_{ij}\bra c_{ij}\ket_{\rm
s}=c\sum_{\bsigma\btau}P(\bsigma)\,P(\btau)\,\Bra \rme^{\beta_f J
\bsigma\cdot\btau}\Ket_J \label{eq:cbar}
\end{equation}
Now $\overline{c}$ represents the average number of connections per
spin in our system. It depends on the replica dimension $n$ via the
scalar spin product and it reduces to $\overline{c}=c$ in the limit
$n\to 0$. In the limit $c\to\infty$ (scaling $J$ as $J/c$ to keep
the local fields in the graph $h_i^{\rm gr}(\bsigma)\equiv \sum_j
c_{ij}J_{ij}\s_j$ of $\mathcal{O}(1)$) we again recover
$\overline{c}=c$ to leading order as found in
\cite{wemmenhove-skantzos-coolen}. Similarly to the above, one also
finds that taking $H_s\to H_s+\lambda\sum_{i<j} c_{ij}J_{ij}$
produces the average bond strength on the graph.
%\begin{equation}
%\overline{J}\equiv \frac{1}{N}\sum_{ij}\bra c_{ij}J_{ij}\ket_{\rm s}
%=c\sum_{\bsigma\btau}P(\bsigma)\,P(\btau)\,\Bra J\, \rme^{\beta_f J
%\bsigma\cdot\btau}\Ket_J
%\end{equation}
As well as being interested in the above average connectivity and
bond strength at total equilibrium, we would also like to
investigate the connectivity structure in more detail. To make
contact with a variety of recent work on complex networks
\cite{doro,newman} we define the degree distribution for our
system
\begin{equation}
\Xi(k)=\lim_{N\to\infty}\frac1N\sum_i\Bra\delta_{k,\sum_jc_{ij}}\Ket_{\rm s}
\end{equation}
Following a calculation similar to that of the free energy in
section \ref{sec:f_extr} one easily finds that
\begin{equation}
\Xi(k)\sim\int
\frac{\rmd\hat{k}}{2\pi}\,\rme^{ik\hat{k}}\sum_{\bsigma\bsigma'}V(\bsigma)U(\bsigma')\,
\exp[c\sum_{\btau}\bra P(\btau)\,\rme^{\beta_f
J\bsigma\cdot\btau-\rmi\hat{k}}\ket_J+\beta_f
J_0\bsigma\cdot\bsigma']
\end{equation}
The above observables are all expressed in terms of the trio of
distributions $P(\bsigma)$, $V(\bsigma)$ and $U(\bsigma)$, taken
at the saddle-point of the free energy (\ref{eq:fs_extr}). To
proceed with a numerical evaluation of the observables one now
needs to specify a form for these.

%%%%%%%%%%%%%%%%%%%%%%%%%%%%%%%%%%%%%%%%%%%%%%%%%%%%%%%%%%%%
\section{Replica symmetry and transfer-matrix diagonalisation}
%%%%%%%%%%%%%%%%%%%%%%%%%%%%%%%%%%%%%%%%%%%%%%%%%%%%%%%%%%%%
\label{sec:RS}

To solve the self-consistent equation (\ref{eq:Ptr}) one is required
to make certain assumptions. Firstly with regards to the form of the
order function $P(\bsigma)$ and the eigenvectors. They represent
different distributions over replicated spins (for any
$n\in\mathbb{R}$). We will consider the simplest possible scenario
in which permutation of spins within different replica groups
$\alpha=1,\ldots,n$ leave the order function invariant (replica
symmetry). This is equivalent to assuming the existence of a single
pure state.

For any natural $n\in\mathbb{N}^+$ it is relatively
straightforward to express these distributions, as their support
is a finite discrete set. However, for the more general case of
$n\in\mathbb{R}$ one has to make an analytic continuation which
leads to more complicated expressions.

For any natural $n$ we can impose replica symmetry by writing for
any arbitrary distribution $X(\bsigma)$
\begin{equation}
X(\bsigma)=\sum_{\ell=0}^n \mathcal{X}(\ell)\
\delta[2\ell-n;\sum_\alpha\s_\alpha]\label{eq:RSN}
\end{equation}
where normalisation of $X(\bsigma)$ requires
$\sum_{\ell=0}^n\mathcal{X}(\ell){n\choose \ell}=1$. On the other
hand, for any real $n$
\begin{equation}
X(\bsigma)=\int \rmd z\ x(z)\
\prod_{\alpha=1}^n\frac{e^{z\s_\alpha}}{[2\cosh(z)]}
\label{eq:RS_nreal}
\end{equation}
where now normalization requires $\int \rmd z\, x(z) = 1$. The above
ans\"atze hold for any distribution $X(\bsigma)$, and in particular
as $X \in \{P,U,V\}$ we define the natural $n$ ans\"atze in terms of
$\{\mathcal{P},\mathcal{U},\mathcal{V}\}$ and for real $n$ in terms
of $\{p,u,v\}$ respectively.

%%%%%%%%%%%%%%%%%%%%%%%%%%%%%%%%%%%%%%%%%%%%%%%%%%%%%%%%%%%%%%%%%%%%%%%%%%%%%%%%%%%
\subsection{Self-consistent order function equations}
%%%%%%%%%%%%%%%%%%%%%%%%%%%%%%%%%%%%%%%%%%%%%%%%%%%%%%%%%%%%%%%%%%%%%%%%%%%%%%%%%%%
\label{sec:Self-consistent_order_function_equations}

Our self-consistent equations for $P,U,V$
(\ref{eq:Ptr}-\ref{eq:T_right}) can now be transformed into
relations between the field distributions
$\{\mathcal{P},\mathcal{U},\mathcal{V}\}$ or $\{p,u,v\}$. Let us
start with the natural $n$ versions. It is convenient to begin by
working out an identity for the replica symmetric form of the
general expression $\sum_{\bsigma} X(\bsigma) F(\bsigma \cdot
\btau)$. We insert the replica symmetric ansatz (\ref{eq:RSN}) for
$X$, use the gauge transformation $\sigma_\alpha \to \sigma_\alpha
\tau_\alpha$ and introduce the representation of unity $1 =
\sum_{k=0}^n \delta[2k-n;\sum_{\alpha} \tau_\alpha]$ which results
in
\begin{equation}
\sum_{\bsigma} X(\bsigma) F(\bsigma \cdot \btau) = \sum_{\ell =
0}^n \sum_{k=0}^n \mathcal{X}(\ell) \delta[2k-n;\sum_{\alpha}
\tau_\alpha] \sum_{\bsigma} \delta[2\ell-n;\sum_\alpha
\sigma_\alpha \tau_\alpha] F(\sum_{\alpha} \sigma_\alpha)
\end{equation}
We now define the set of replica indices $S = \{\alpha \in
\{1,\ldots,n\} : \tau_\alpha = 1\}$ and its complement
$\overline{S}=\{\alpha \in \{1,\ldots,n\} : \tau_\alpha = -1\}$
which allows us to write $\sum_\alpha \tau_\alpha \sigma_\alpha =
\sum_{\alpha \in S} \sigma_\alpha - \sum_{\alpha \in \overline{S}}
\sigma_\alpha$. Isolating these last two summations via the unities
$1 = \sum_{k_1 = 0}^k \delta[2k_1 - k;\sum_{\alpha \in S}
\sigma_\alpha]$ and $1 = \sum_{k_2 = 0}^{n-k} \delta[2k_2 +k -n;
\sum_{\alpha \in \overline{S}} \sigma_\alpha]$ and using the general
identity $\sum_{\sigma_1\ldots\sigma_p} \delta[2q-p;\sum_{\alpha =
1}^p \sigma_\alpha] = {p \choose q}$, we obtain
\begin{eqnarray}
\lefteqn{\sum_\bsigma X(\bsigma) F(\bsigma \cdot \btau) =
\sum_{\ell,k = 0}^n \sum_{k_1 = 0}^k \sum_{k_2 = 0}^{n-k}
\mathcal{X}(\ell)\,\delta[2k-n;\sum_\alpha \tau_\alpha]}\nonumber\\
&& \times\, \delta[\ell + k + k_2 - k_1 - n;0]\, F(2(k_1 + k_2) -
n)\, {k \choose k_1} {n-k \choose k_2}
\end{eqnarray}
Using the above identity (and very similar manipulations) we can
write our self-consistent equations as
\begin{eqnarray}
\mathcal{U}(\ell)
&=&
\lambda_0^{-1}(n)\, \exp[c A_{P}(\ell, J)]\,
A_{U}(\ell, J_0)
\\
\mathcal{V}(\ell) &=& \lambda_0^{-1}(n)\, \sum_{j = 0}^n \sum_{k_1 =
0}^j \sum_{k_2 = 0}^{n-j}\, \mathcal{V}(j)\, \rme^{c A_P(j,J) +
\beta_fJ_0[2(k_1 + k_2) - n]} \,\delta[\ell +j + k_2 - k_1 - n; 0]
\\
\mathcal{P}(\ell)
&=&
\frac{\mathcal{U}(\ell)\, \mathcal{V}(\ell)}{\sum_{\ell = 0}^n {n
\choose \ell}\,\mathcal{U}(\ell)\, \mathcal{V}(\ell)}
\end{eqnarray}
The largest eigenvalue $\lambda_0(n)$ follows from the above by
utilizing the normalization condition $\sum_\ell {n\choose \ell}
\mathcal{U}(\ell)=1$. We have introduced the convenient shorthand,
\begin{eqnarray}
A_X(\ell,J) = \sum_{i = 0}^n \sum_{j = 0}^\ell \sum_{k = 0}^{n -
\ell} \mathcal{X}(i)\ \delta[i+k+\ell-j-n;0]\ {\ell \choose j} {n-
\ell \choose k} \bra \rme^{\beta_f J[2(j+k)-n]} \ket_J
\end{eqnarray}
for $X\in\{P,U,V\}$ and
$\mathcal{X}\in\{\mathcal{P},\mathcal{U},\mathcal{V}\} $
respectively.

 Let us now turn to the more general case of
$n\in\mathbb{R}$. Firstly, a Taylor expansion of the transfer matrix
elements (\ref{eq:transfer_matrix}) into a series of exponentials
and insertion of the replica symmetric ansatz (\ref{eq:RS_nreal})
leads to
\begin{equation}
T_{\bsigma,\bsigma'}[P]= \rme^{\beta_f J_0\bsigma\cdot\bsigma'}\Bra
\rme^{\beta_f\theta\sum_{\alpha}\s_\alpha} \Ket_\theta
\end{equation}
with $\bra\cdots\ket_\theta$ representing averages over the measure
\begin{equation}
M(\theta|n)= \sum_{k\geq 0} \frac{\rme^{-c}c^k}{k!}\Bra \int[\prod_{l\leq k}
\frac{\rmd h_l\,p(h_l)}{[2\cosh(h_l)]^n} ]\
\rme^{n\sum_lB(J_l,h_l)}\
\delta[\theta-\sum_{l\leq k}A(J_l,h_l)]\Ket_{\{J_l\}} \label{eq:M}
\end{equation}
and where we introduced the functions
\begin{eqnarray}
A(J,x) & = & {\rm atanh}(\tanh(\beta_f J)\tanh(x))
\\
B(J,x) & = & \frac{1}{2}\log\left[4\cosh[\beta_f J+x]\cosh[\beta_f
J-x]\right]
\end{eqnarray}
The first of the above functions can be identified as a `message'
(or effective field) passed during belief propagation while the
latter is related to the free energy shifts which occur during an
iteration \cite{mezard-parisi87,mezard-parisi01}. For $n \to 0$ and
within replica symmetry this second term does not contribute
although in the more general case of $n>0$ it will play an important
role. Following the belief propagation picture, one can also relate
(\ref{eq:M}) to a weighted measure over the messages coming from the
long range bonds. Performing the spin summations in
(\ref{eq:T_left}-\ref{eq:T_right}) using the ansatz
(\ref{eq:RS_nreal}) and requiring the resulting expression to have
the eigenvector form leads to
\begin{eqnarray}
\lambda_0(n)\ u(x|n) & = & \int \rmd x'\ u(x'|n)\
\frac{\cosh^n(x)}{\cosh^n(x^\prime)} \Bra \rme^{n
B(J_0,x')}\delta[x-\theta-A(J_0,x')]\Ket_\theta \label{eq:u}
\\
\lambda_0(n)\ v(y|n) & = & \int \rmd y'\ v(y'|n)\
\frac{\cosh^n(y)}{\cosh^n(y^\prime)} \Bra \rme^{n
B(J_0,y'+\theta)}\delta[y-A(J_0,y'+\theta)]\Ket_\theta \label{eq:v}
\end{eqnarray}
so that the largest eigenvalue follows from the above by simple
integration. To close the above equations we also need to derive an
expression for the function $p(h)$. The starting point for this is
equation (\ref{eq:Ptr}). Rewriting the traces in terms of the
eigenvectors and substituting our ansatz (\ref{eq:RS_nreal}) results
in
\begin{equation}
p(h) = \frac{\int  \rmd x\, \rmd y\ u(x)\, v(y)\ \left\{
\frac{\cosh(h)}{2\cosh(x)\cosh(y)} \right\}^n\
  \delta[h -  (x + y)]}{\int \rmd x\, \rmd y\ u(x)\,
  v(y)\left\{\frac{\cosh(x+y)}{2\cosh(x)\cosh(y)}\right\}^n} \label{eq:ph}
\end{equation}
The coupled set of equations (\ref{eq:u},\ref{eq:v},\ref{eq:ph}) are
to be solved self-consistently. This has a clear interpretation in
terms of message-passing algorithms: $p(h)$ gives the distribution
of messages passed along long-range short-cuts, whereas, $u(x)$ and
$v(y)$ that of messages passed along the chain (from the left- and
right- neighbors). This inspires a solution using a population
dynamics methodology \cite{mezard-parisi01}; the main difference is
that one is now required to weight the averages of the field
distributions by an $n$-dependent factor. In practice, expressions
of the form $\phi(x')=\int \rmd x\, \phi(x)\,
w(x)\,\,\delta[x'-g(x)]$ (for some arbitrary probability density
$\phi(x)$, weight $w(x)$ and updating function $g(x)$) are solved by
sampling values of $x$ from the density $\phi(x)$ and updating
$x'\to g(x)$ with weight $w(x)$. To interpret this weighting term,
one can write $w(x)=\lfloor w(x)\rfloor+p$, where, $\lfloor
w(x)\rfloor$ is the integer part of $w(x)$, and $p$ is the
fractional part. At each iteration step we replace $\lfloor w(x)
\rfloor$ of the population members with $x^\prime$ and a further
member with probability $p$.

%%%%%%%%%%%%%%%%%%%%%%%%%%%%%%%%%%%%%%%%%%%%%%%%%%%%%%%%%%%%%%%%%%%%%%%%%%%%%%%%%
\subsection{Observables within replica symmetry}
%%%%%%%%%%%%%%%%%%%%%%%%%%%%%%%%%%%%%%%%%%%%%%%%%%%%%%%%%%%%%%%%%%%%%%%%%%%%%%%%%
\label{sec:Observables within replica symmetry}

\begin{figure}[t]
\setlength{\unitlength}{0.1cm}
\begin{picture}(200,60)
\put(30,5){\includegraphics[height=6.5cm,width=8.5cm]{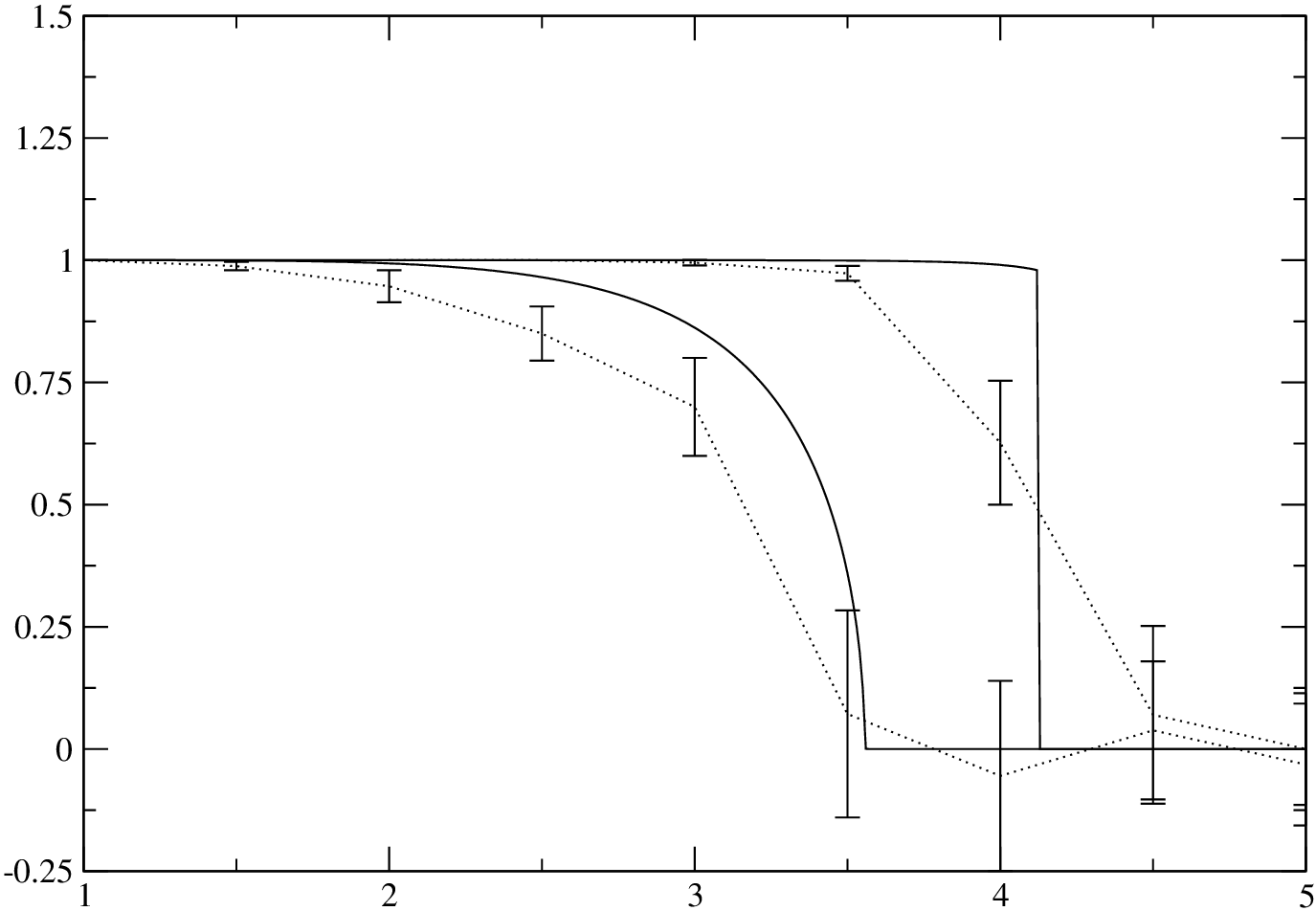}}
\put(70,0){$T_f/J$} \put(25,33){$m$}
\end{picture}
\caption{\label{fig:mag} We plot the magnetization $m$ as a function
of temperature $T_f$ for $n=1$ and $n=5$. The solid lines are the
theoretical predictions while the dotted lines are a guide for the
eye joining the markers with error bars which come from simulations.
We have $c =2$ and $J_0 = J_{ij} = 1\ \forall i,j$. The simulations
were done via Monte Carlo Glauber dynamics (see text for details) on
$N = 200$ spins. Despite the small system size they seem to be in
reasonable agreement with the theory.}
\end{figure}

To calculate the magnetization (\ref{eq:m_def}) and spin glass
(\ref{eq:q_def}) order parameter within our system, we substitute
the replica symmetric ansatz for $P(\bsigma)$ (\ref{eq:RSN}) into
their definitions, which together with a minor rearrangement gives
for integer $n$
\begin{eqnarray}
m &=& \sum_{\ell = 1}^n \mathcal{P}(\ell) {n-1 \choose \ell - 1} -
\sum_{\ell = 0}^{n-1} \mathcal{P}(\ell) {n-1 \choose \ell} \\
q &=& \sum_{\ell = 2}^n \mathcal{P}(\ell) {n-2 \choose \ell-2} +
\sum_{\ell = 0}^{n-2} \mathcal{P}(\ell) {n-2 \choose \ell} -
\sum_{\ell = 1}^{n-1} \mathcal{P}(\ell) {n-2 \choose \ell-1}
\end{eqnarray}
while for real $n$ the expressions become:
\begin{eqnarray}
m&=&\int \rmd h\ p(h)\ \tanh(h) \label{eq:m}\\
q&=&\int \rmd h\ p(h)\ \tanh^2(h) \label{eq:q}
\end{eqnarray}
So given $\mathcal{P}$ or $p$ from the self-consistent equations,
for either integer or real $n$, we may evaluate these order
parameters. In figure \ref{fig:mag} we plot the magnetization for
two different values of $n$, and compare our results against
simulation experiments. More details on the simulations are given in
section \ref{sec:simul}, but we note here that these experiments are
particularly time consuming due to the coupled dynamics, so that,
only modest system sizes are allowed for reasonable CPU cost. Within
these constrains we feel that the agreement is reasonable.

Evaluating the fraction of energetically non-optimal bonds $\psi$ is
slightly more involved. We use the replica symmetric representation
of the transfer matrix which for integer $n$ reads
\begin{eqnarray}
T_{\bsigma,\bsigma^\prime}[P] = \sum_{k = 0}^n
\delta(2k-n;\sum_\alpha \sigma^\alpha)\ \rme^{\beta_f J_0 \bsigma
\cdotp \bsigma^\prime + c A_P(k,J)}
\end{eqnarray}
Then, after some combinatorial work, $\psi$ is found to be given by
\begin{eqnarray}
\psi &=& D_1\sum_{i = 1}^n  \sum_{j= 0}^{n-1}  \sum_{k=0}^{i-1}
\rme^{\beta_f J_0(2k - j -1)} {n-1 \choose i-1} {i-1 \choose k}{n-i
\choose j-k}\Bigg\{\mathcal{V}(i)\mathcal{U}(j) \rme^{cA_P(i,J)}
+\mathcal{V}(j) \mathcal{U}(i) \rme^{c A_P(j, J)} \Bigg\}\nonumber
\\
& &
+ D_2 2p\sum_{i=1}^n \sum_{j = 0}^{n-1} \sum_{k = 0}^{i-1}
\mathcal{P}(i) \mathcal{P}(j) {n-1 \choose i-1} {i-1 \choose k} {n-i
\choose j-k} \rme^{\beta_f J (n + 2(2k - i - j))}\nonumber\\
& &
+ D_2 (1-p) \sum_{i = 1}^n \sum_{j = 1}^n \sum_{k = 0}^{i-1}
\mathcal{P}(i) \mathcal{P}(j) {n-1 \choose i-1} {i-1 \choose k} {n -
i \choose j-k-1} \rme^{\beta_f J(n + 2(2k +1 - i-j))}\nonumber\\
& &
+ D_2 (1-p) \sum_{i = 0}^{n-1} \sum_{j = 0}^{n-1} \sum_{k = 0}^{i}
\mathcal{P}(i) \mathcal{P}(j) {n-1 \choose i}  {i \choose k} {n-i-1
\choose j -k} \rme^{\beta_f J(n + 2(2k - i-j))}
\end{eqnarray}
For real $n$ the replica symmetric transfer matrix is given by,
\begin{eqnarray}
T_{\bsigma, \bsigma^\prime}[P] = \rme^{\beta_f J_0 \bsigma \cdotp
\bsigma^\prime} \int \rmd \theta M(\theta)\ \rme^{\beta_f \theta
\sum_\alpha \sigma_\alpha}
\end{eqnarray}
leading to
\begin{eqnarray}
\psi &=&
D_1 \int \rmd x \rmd y \rmd \theta\ u(x) v(y)
M(\theta)\ \frac{ \rme^{-\beta_f J_0}\, 2\cosh(x + \theta - y)}{
\big[\rme^y 2\cosh (x + \theta + \beta_f J_0) + \rme^{-y} 2\cosh(x +
\theta - \beta_f
J_0)\big]^{n-1}}\nonumber
\\
& & +D_2 \int \rmd h_1 \rmd h_2 \ p(h_1) p(h_2)\ \rme^{-\beta_f J}\,
\frac{Q(J) 2\cosh(h_1 + h_2) + Q(-J) 2\cosh(h_1 - h_2)}{[\rme^{h_2}
2\cosh(h + \beta_f J) + \rme^{-h_2} 2\cosh(h - \beta_f J)]^{n-1}}
\end{eqnarray}
where we define
\begin{eqnarray} Q(J^\prime) = r \delta_{J, J^\prime} +  (1-r)\delta_{-J, J^\prime}
\qquad r= \frac{\rme^{K_p J}}{2\cosh(K_p J)}
\end{eqnarray}

\begin{figure}[t]
\setlength{\unitlength}{0.1cm}
\begin{picture}(200,60)
\put(30,5){\includegraphics[height=6.5cm,width=8.5cm]{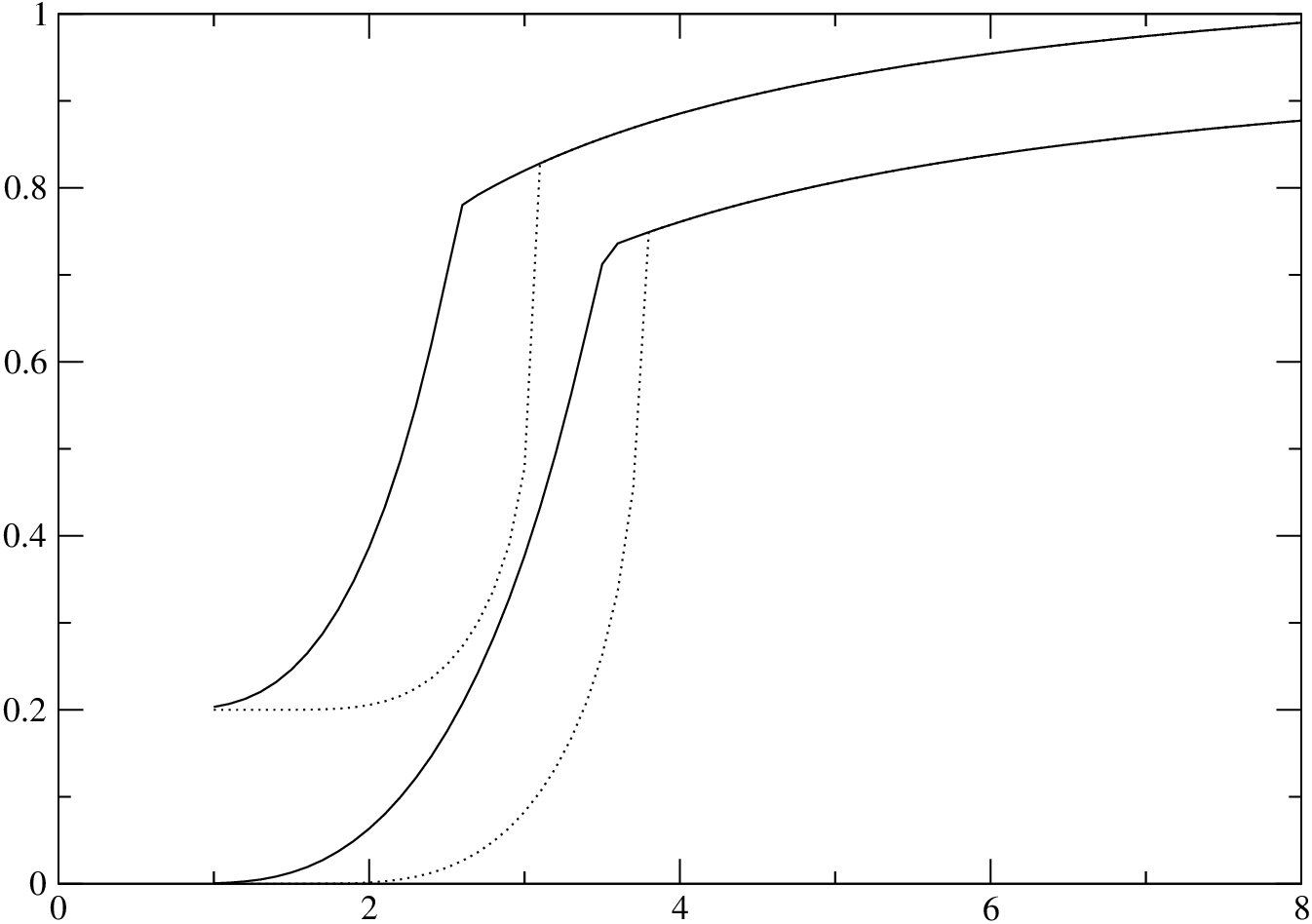}}
\put(70,0){$T_f/J$} \put(25,33){$\psi$}
\end{picture}
\caption{\label{fig:frust} We plot the fraction of misaligned bonds
$\psi$ against the re-scaled temperature $T_f/J$ for $c = 2$ and
$J_0 = 1$. The solid lines are for $n = 1$ while the dotted lines
are for $n = 3$. The upper pair of lines are for $r= 0.8$ while the
lower pair are for the ferromagnet, $r = 1$. We see that in the
ordered phase, increasing $n$ allows the system to optimize the
bonds energetically.}
\end{figure}

We plot $\psi$ for a few sets of parameters in figure
$\ref{fig:frust}$. As the fast temperature goes to $\infty$ we have
$\psi \to 1$, i.e. exactly half of the bonds at any point in time
are energetically optimal, so in the high temperature phase the
ordering is non-existent (as one would expect). We also see that
increasing $n$ leads to better levels of optimization, again what we
would expect but it is possible to quantify it here. Decreasing $r$
and hence increasing the disorder, makes it harder for the spins to
energetically optimize themselves, although at low temperatures, due
to the condensation phenomena in the bonds (see below), the
magnetization will increase to 1. With all spins aligned, the
fraction of energetically non-optimal bonds becomes exactly the
fraction of bonds with $J < 0$. We see this in figure
\ref{fig:frust}, as $T_f \to 0$, $\psi \to 1-r$. This would not be
the case for the pure spin-glass, $r = 0.5$.

We now turn our attention to the graph observables. We first focus
on the average connectivity which is expressed as
\begin{eqnarray}
\overline{c} = c \sum_k \mathcal{P}(k) A_P(k;J) {n \choose k}
\end{eqnarray}
for integer $n$ and
\begin{eqnarray}
\overline{c} = c \int \rmd h_1\, \rmd h_2\ p(h_1)\, p(h_2)\
\left\{\cosh(\beta_f J) + \tanh(h_1) \tanh(h_2) \sinh(\beta_f J)
\right\}^n
\end{eqnarray}
for real $n$.

\begin{figure}[t]
\setlength{\unitlength}{0.1cm}
\begin{picture}(200,60)
\put(30,5){\includegraphics[height=6.5cm,width=8.5cm]{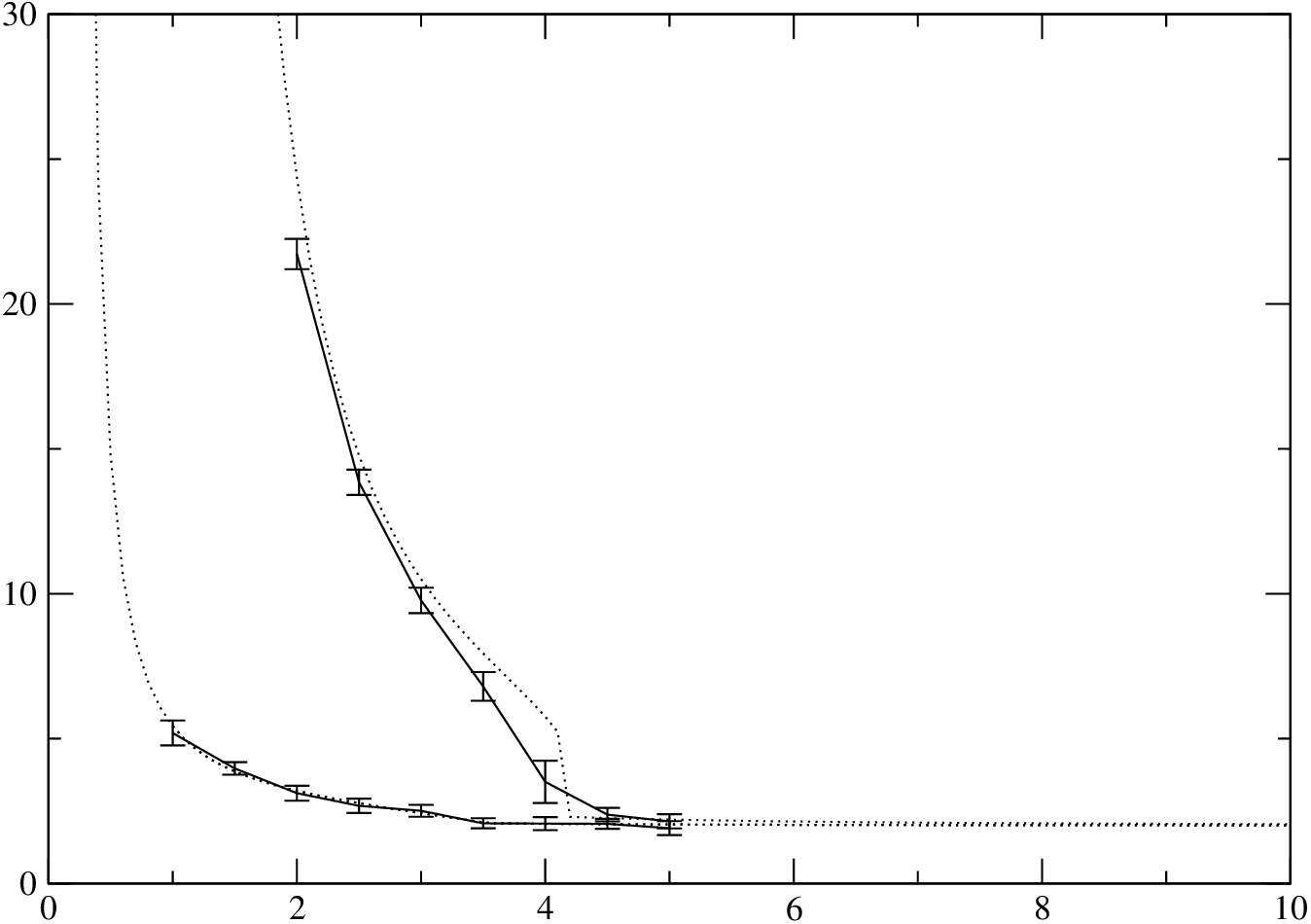}}
\put(70,0){$T_f/J$} \put(25,33){$\overline{c}$}
\end{picture}
\caption{\label{fig:caver} We plot the average number of bonds
$\overline{c}$ against temperature for $J = J_0 = 1$ and $c = 2$ in
the ferromagnet. The higher line with the first order transition is
for $n=5$ while the lower line is for $n=1$. The dotted lines are
the theoretical predictions while the solid line is a guide for the
eye linking the error bars which are measurements from simulation
experiments. The agreement is reasonable although, as in fig.\@ \ref{fig:mag}, we
find that the sharp transition is smeared out due to the small
system size for our simulations ($N$ = 200 spins).}
\end{figure}

\begin{figure}[h]
\setlength{\unitlength}{0.1cm}
\begin{picture}(200,70)
\put(30,5){\includegraphics[height=6.5cm,width=8.5cm]{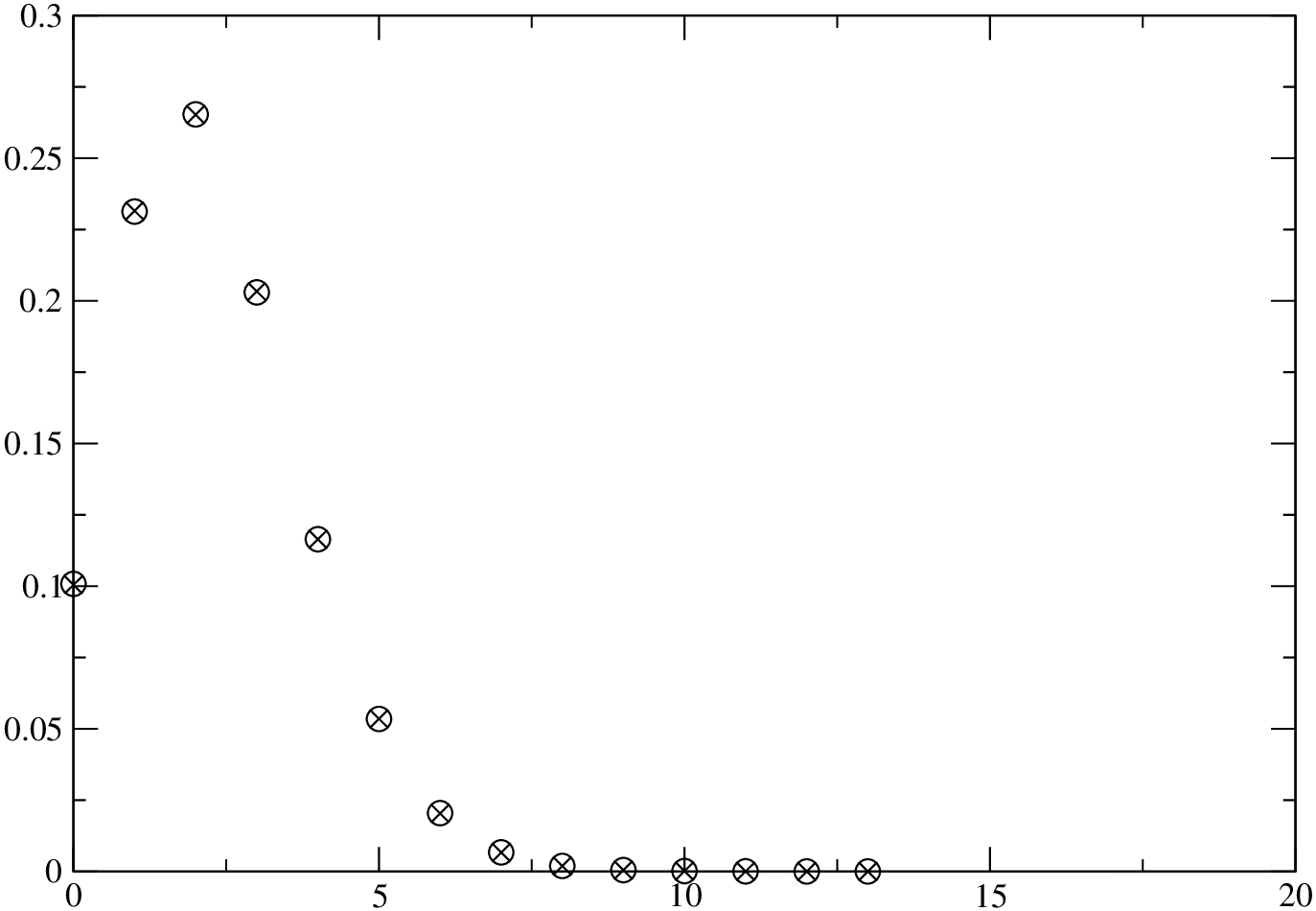}}
\put(70,0){$k$} \put(25,33){$p_k$}
\end{picture}
\caption{\label{fig:degreedist} We plot the probability that a given
node has degree $k$, $p_k$ against $k$ for $n = 0.5$, $r = 0.6$,
$T_f = J = J_0 = 1$ and $c = 2$ within the spin-glass regime where
other observable values are $m = 0$ and $q \approx 0.581$. The true
degree distribution is given by crosses, for comparison we have also
given the Poisson distribution with the same value of $\overline{c}$
with circles. Although there are differences between the two for
this set of parameters, the difference is very small.}
\end{figure}

If figure $\ref{fig:caver}$ we plot the average number of bonds
$\overline{c}$. At low temperatures (the specific temperature
depends on other parameters) the average connectivity increases
sharply. This is due to ordering within the spin system leading to
increased energetic gain by adding connections. Higher values of
$n$, for a given $T_f$, means a lower value of $T_s$ and hence the
connectivity variables will be governed more strictly by the free
energy of the fast system, which is minimized by high connectivity
configurations.

We also looked at the full degree distribution, which is given up to
normalization constants for integer $n$ by
\begin{eqnarray}
\Xi(k) \sim \frac{c^k}{k!} \sum_{i = 0}^n {n \choose i}\
\mathcal{V}(i)\ \mathcal{U}(i)\ A_P^k(i, J)
\end{eqnarray}
and for real $n$ by,
\begin{eqnarray}
\Xi(k) \sim \frac{c^k}{k!} \int \prod_{\ell \leq k} \left\{
\frac{\rmd h_\ell \rmd J_\ell p(h_\ell)
Q(J_\ell)}{[2\cosh(h_\ell)]^n} \right\} \frac{\rmd x \rmd y u(x)
v(y)}{[4\cosh(x)\cosh(y)]^n} \rme^{n \sum_\ell B(J_\ell, h_\ell) + n
B(J_0, x)}
\\
\times\
\left\{2\cosh\left[y + A(J_0, x) + \sum_\ell A(J_\ell, h_\ell)
 \right] \right\}\nonumber
\end{eqnarray}
A typical example of this degree distribution is given in figure
$\ref{fig:degreedist}$. What is particularly interesting is that
although the degree distribution is in principle free to take on any
form it keeps very close to that of the Poisson degree distribution
with mean $\overline{c}$. In fact in the paramagnetic phase, we know
that $P(\bsigma) = 2^{-n}$ and thus we find $\overline{c}$ exactly
from (\ref{eq:cbar}) without invoking replica symmetry, namely
$\overline{c}_{PM} = c \cosh^n(\beta J)$, which is independent of
$r$ since cosh is an even function. Thus the average degree is
independent of the bond disorder (in this model) in the paramagnetic
phase. Here, the degree distribution also scales linearly with $c$.
By using the fact that in the paramagnetic phase we also have
$U(\bsigma) = V(\bsigma) = 2^{-n}$ we can also see that $\Xi(k) =
\rme^{-\overline{c}_{PM}}\overline{c}_{PM}^k/k!$, i.e. the degree
distribution is exactly Poisson. We also find exact results in the
fully ferromagnetic phase where $P(\bsigma) = U(\bsigma) =
V(\bsigma) = \prod_{\alpha} \delta_{\sigma^\alpha, \sigma^1}$. There
$\overline{c}_{FM} = c \bra \rme^{\beta_f n J} \ket_J$ and $\Xi(k) =
\rme^{-\overline{c}_{FM}}\overline{c}_{FM}^k/k!$. In both these
cases the degree distribution is exactly Poisson. We can understand
this, since in both phases there is no energetic gain in having any
particular $c_{ij} = 1$, since it will not affect the spin
distribution (they are either all set to be aligned, or fully
random) and thus the degree distribution will be the maximum entropy
one, i.e. Poisson. It is also clear that there are a range of
ordered states between the two extremes above, and that we cannot
say anything further analytically about the degree distribution
there. However, we may evaluate our order parameter equations
numerically, and we find that although the degree distribution is
not Poisson, it is very close, as shown in figure
\ref{fig:degreedist}. It was not obvious that this should be the
case, and indeed, the increased critical temperature for a scale
free degree distribution would have suggested that this could be
optimal, since it increases ordering, but it transpires that this is
not the case here.

%%%%%%%%%%%%%%%%%%%%%%%%%%%%%%%%%%%%%%%%%%%%%%%%%%%%%%%%%
\subsection{Phase diagrams}
%%%%%%%%%%%%%%%%%%%%%%%%%%%%%%%%%%%%%%%%%%%%%%%%%%%%%%%%%
\label{sec:phase_diagrams}

\begin{figure}[t]
\setlength{\unitlength}{0.1cm}
\begin{picture}(200,70)
\put(10,5){\includegraphics[height=5.5cm,width=7.0cm]{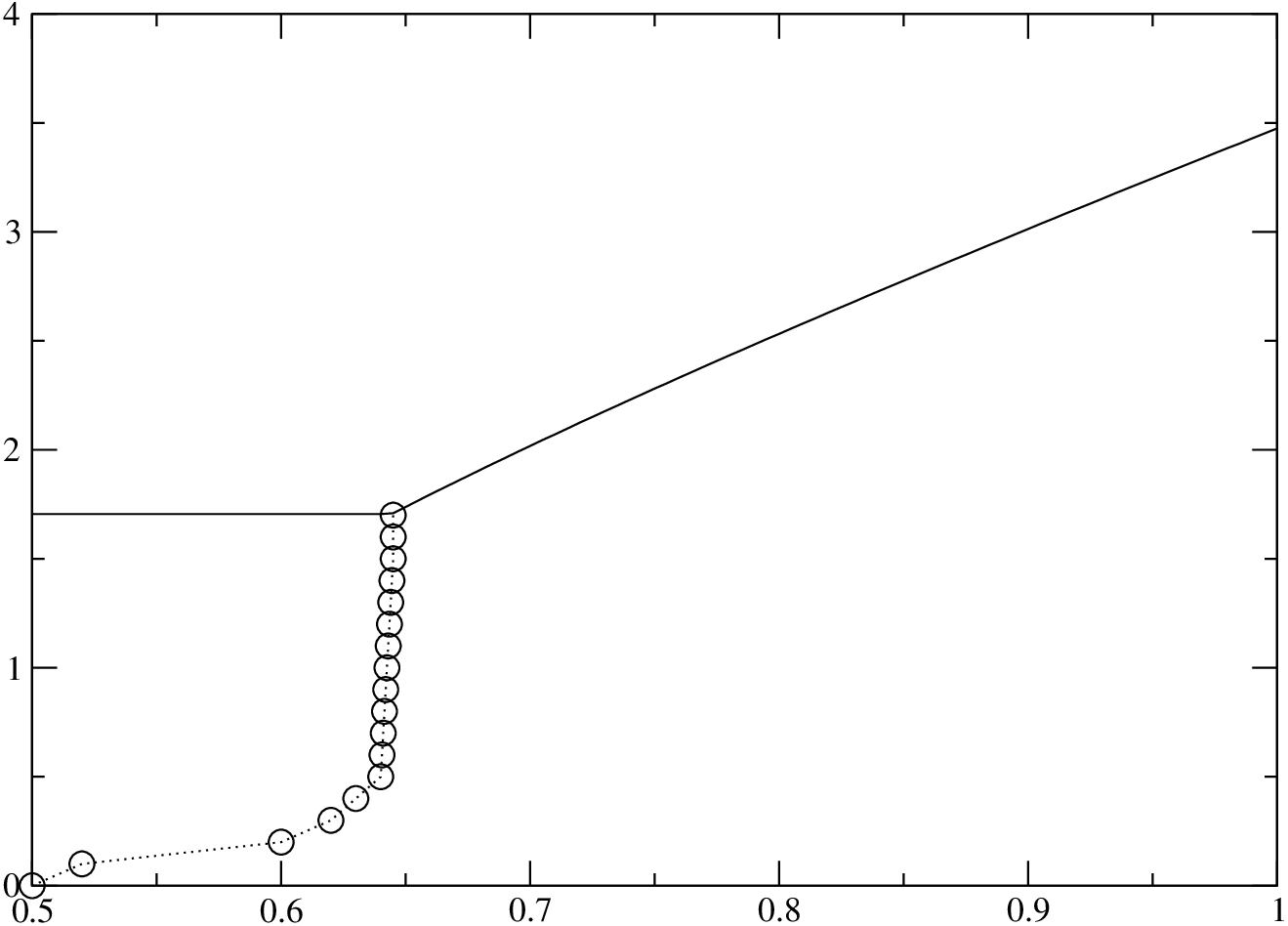}}
\put(90,5){\includegraphics[height=5.5cm,width=7.0cm]{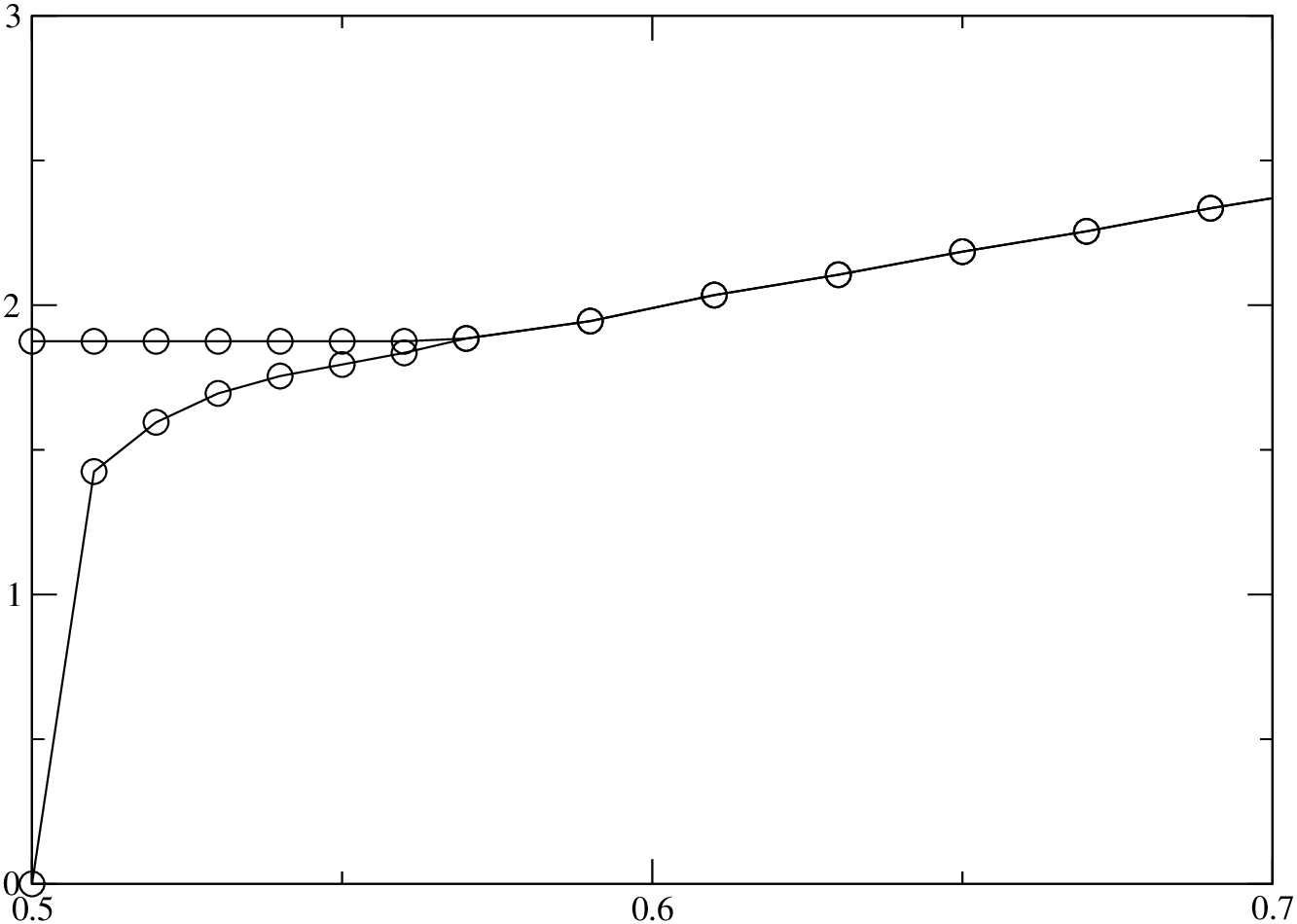}}
\put(50,0){$p$} \put(0,33){$T_f$} \put(120, 0){$p$}
\put(30,45){P}\put(120,50){P} \put(20,20){SG}\put(92,34){SG}
\put(55,20){F}\put(120,20){F}
\end{picture}
\caption{\label{fig:phased} We plot the phase diagrams for $c=2$ and
$J = J_0 = 1$. The left figure is for $n = 0.1$, where the solid
lines are given by the bifurcation conditions (\ref{eq:PtoF},\ref{eq:PtoSG})
while the markers come
from solving the order parameter equations numerically and the
dotted line linking markers is a guide to the eye. The right figure
is for $n = 2$ and all lines are linking markers which come from solving the
order parameter equations numerically. The P$\to$F and P$\to$SG transitions are here first-order. For larger values of $n$ we
see that the links are better able to align to increase order; firstly, the transition temperature from the paramagnetic
phase is higher and secondly, the size of the spin glass phase is significantly smaller.}
\end{figure}
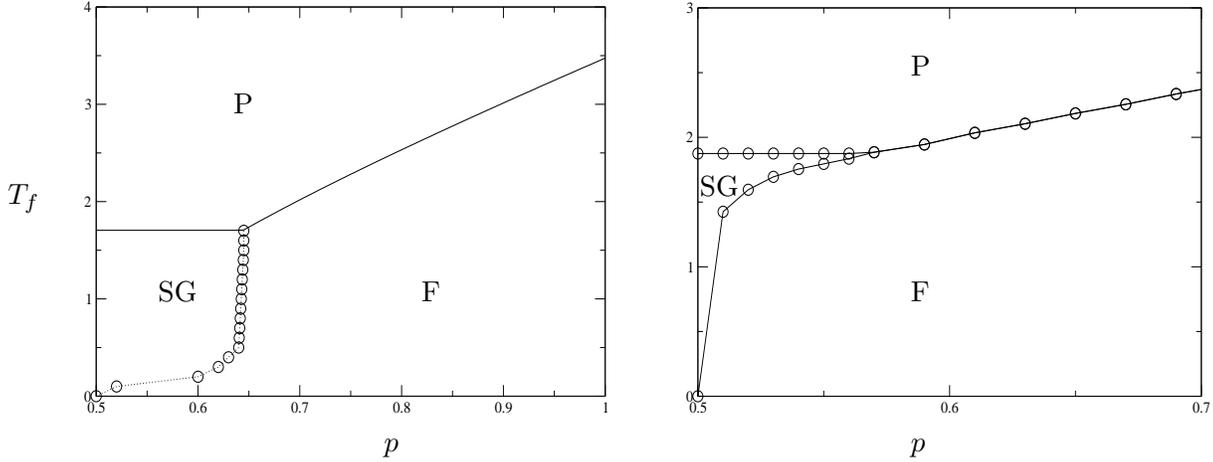

Having derived the main equations from which our observables follow,
we can now proceed to the evaluation of the transition lines in our
phase diagram numerically and via a bifurcation analysis. Firstly,
we see that the state $p(x)=u(x)=v(x)=\delta(x)$ always solves
equations (\ref{eq:u},\ref{eq:v},\ref{eq:ph}), giving $m=q=0$, for
all temperatures. We can therefore associate this state with the
high-temperature (paramagnetic) solution. To examine continuous
bifurcations away from this solution we assume that close to the
transition the fields are small and that the paramagnetic
$\delta$-distributions evolve to either distributions of small,
non-zero mean (in leading order) marking the
paramagnetic/ferromagnetic transition or to distributions of small,
non-zero variance (again in leading order) marking the
paramagnetic/spin-glass transition. With these considerations in
mind we define the moments $\overline{h^\ell}=\int \rmd
h\,p(h)\,h^\ell=\order(\epsilon ^\ell)$ for some $0<\epsilon\ll 1$
(and similarly for $\overline{x^\ell}=\int \rmd x\,u(x)\,x^\ell$ and
$\overline{y^\ell}=\int \rmd y\,v(y)\,y^\ell$). We assume that there is no first order transition. Then, expanding equations
(\ref{eq:u},\ref{eq:v},\ref{eq:ph}) for small values of fields and
using $\overline{h}=\overline{x}+\overline{y}$ and
$\overline{h^2}=\overline{x^2}+\overline{y^2}$ which follows from
(\ref{eq:ph}) we arrive at paramagnetic/ferromagnetic and
paramagnetic/spin-glass transition lines:
\begin{eqnarray}
{\rm P \to F}: \quad & & 1 = c \Bra \sinh(\beta_f J) \cosh^{n-1}(\beta_f
J)\Ket_J \rme^{2 \beta_f J_0} \label{eq:PtoF}\\
{\rm P \to SG}:\quad && 1 = c \Bra \sinh^2(\beta_f J) \cosh^{n-2}(\beta_f J)
\Ket_J \cosh(2\beta_f J_0)\label{eq:PtoSG}
\end{eqnarray}
These equations reduce to those found in \cite{guzaiSW} in the limit
$n \to 0$, recovering the small-world bifurcations. The correspondence is exact
if we identify the paramagnetic mean
connectivity here with $c$. It also reduces to those
of \cite{wemmenhove-skantzos} for $J_0 = 0$ and $Q(J^\prime) =
\delta(J^\prime - J)$, where the Hopfield model on a dynamic random
graph was studied, if in the latter only a single pattern is stored
(in this scenario the Hopfield model becomes equivalent to a
ferromagnet with a different gauge). It is well known for these
models \cite{wemmenhove-skantzos} that as $n$ increases, the
transitions are increasingly likely to be first order. Thus to
produce phase diagrams of the system, as well as looking at the
bifurcation lines given by the above we also solved the full
equations numerically. The results are shown in figure
\ref{fig:phased} where we see that increasing $n$ decreases the size
of the spin glass phase, which we expect is due to the increased
cooperativity.

\subsection{Simulations}
\label{sec:simul}
 In order to check the validity of our theoretical
work we performed numerical simulations of this model. To do this we
need to introduce a dynamical process on both the spins and the
graph which will converge to an equilibrium distribution described
by their respective Hamiltonians. One way to do this is via Glauber
dynamics \cite{Glauber63}, the dynamics then automatically obey
detailed balance. The transition rates between a given state and
another state with a single "spin" flip (where we take spin in the
broader sense to include the binary variables $\{c_{ij}\}$ and
$\{J_{ij}\}$ as well as the more familiar $\{\sigma_i\}$) is
determined by half the energy difference (or local field) between
the two states. Defining general spin flip operators via $F_{ij}^c
\Phi(c_{11},c_{12},\ldots,c_{NN}) =
\Phi(c_{11},c_{12},\ldots,-c_{ij},\ldots,c_{NN})$ and similarly for
$F_{ij}^J$ the Glauber rates can be written as
\begin{eqnarray}
W[F_{ij}^c \mathbf{c}, \mathbf{c}] &=& \frac12 \left\{1 -
\tanh\left[\frac{2c_{ij}-1}{2} \log\frac{c}{N} - \frac{n}{2} \log
\bra \rme^{-\beta_f(2c_{ij} - 1) \sigma_i J_{ij} \sigma_j} \ket
\right] \right\} \\
W[F_{ij}^J \mathbf{J}, \mathbf{J}] &=& \frac12\left\{1 -
\tanh\left[J_{ij}K_p -\frac{n}{2} \log\bra \rme^{-2\beta_f \sigma_i
c_{ij} J_{ij} \sigma_j } \ket \right] \right\}
\end{eqnarray}
where the angular brackets denote averages over the fast process for
the given realization of the graph and bonds.

The nature of the coupled dynamics means that for each change to the
graph (the slow dynamics) one must re-equilibrate the spins, measure
the averages as required in the above equation and subsequently change
the graph configuration again. Thus the computation effort
required to equilibrate the slow system is very large compared to
simulations on a given, fixed, graph. In particular, for strongly disordered
graphs, where changing a single bond is expected to seriously alter
the free-energy surface, it is very difficult to obtain reasonable
statistics. Instead, we have focused our efforts on the simpler case
of purely ferromagnetic bonds. This means that after changing a
given bond, the new equilibrium distribution is expected to be very
close to the old one and also the equilibration times will be
shorter in general. We have performed simulations on systems with $N
= 200$, and in figures \ref{fig:mag} and \ref{fig:caver} we compare
the results with our theoretical predictions. Due to the small
system size we must expect that there are both persistent errors due
to the relatively small system size, smearing of all phase
transitions and large error bars on any given measurement. Bearing
this all in mind we feel that the results, particularly for the
average connectivity, clearly support the theory.

\section{Conclusions}

The study of complex networks has recently become a very popular
field due to their ubiquity in nature, technology and social
interaction, where these fields are meant in a broad sense. While
the statistical structure characterizing real world networks (path
lengths, degree distributions,$\ldots$) and models that recreate
these properties have been extensively studied from experimental
measurements on real world systems, through numerical simulations
and theory, understanding the behavior of networked systems based on
local rules (dynamics) is still a relatively unexplored area
\cite{newman}. We have presented a solvable model that examines a
spin system on a small world graph with which we have probed
cooperative behavior of the entire system (both of the graph and the
spins). To overcome the theoretical challenge of systems evolving on
disparate timescales we have focussed on the adiabatic limit; the
graph evolves infinitely slowly relative to the spin variables. This
allows us to treat the model using the well developed thermodynamics
of replica theory, rather than having to treat the dynamics
explicitly. The advantage of this approach is twofold. Firstly the
results are exact in the thermodynamic limit in the region where
replica symmetry is stable. Although we have not examined replica
symmetry breaking experience suggests that this would only occur for
$n < 1$, at low temperatures (high values of $\beta_f$) and for some
critical amount of disorder in the bonds $\{J_{ij}\}$. The second
benefit of this approach is related to the relative simplicity of
our present approach. We do not specify in advance the dynamics of
the graph, but instead only describe it through its equilibrium
energy function. Thus the resulting graph structure becomes an
observable itself, rather than an object which is fixed from the
start. Indeed, naive intuition may suggest that the optimal
structure could have been scale free, so that ordering in the spins
would have occurred at a higher temperature. It turns out that this
was not the case, apparently due to entropic reasons.

%It is quite conceivable to imagine a host of
%generalizations to the above approach, e.g.\@ consider more realistic energy functions or
%modifying the slow Hamiltonian as to promote a different statistics
%on the graph statistics at total equilibrium.

\begin{acknowledgements}
NSS wishes to warmly thank RIKEN Brain Science Institute for their
kind hospitality during the final stages of this work and the Fund for Scientific
Research Flanders-Belgium. We thank A C C Coolen, I P\'erez Castillo and B Wemmenhove
for illuminating discussions.
\end{acknowledgements}

\end{document}